# Asymmetrical Order in Wall-Bounded Turbulent Flows

T.-W. Lee

Mechanical and Aerospace Engineering, SEMTE

Arizona State University, Tempe, AZ 85287

**Abstract**

Scaling of turbulent wall-bounded flows is revealed in the gradient structures, for each of the Reynolds stress components. Within the "dissipation" structure, an asymmetrical order exists, that we can deploy to unify the scaling and transport dynamics within and across these flows. There are subtle differences in the outer boundary conditions between channel and flat-plate boundary-layer flows, which modify the turbulence structure far from the wall. The self-similarity exhibited in the dissipation structures and corresponding transport dynamics establish capabilities and encompassing knowledge of wall-bounded turbulent flows.

**INTRODUCTION**

Scaling in turbulent flows provides a format upon which the observed structural characteristics can be unified. In turbulent jets, as an example, the radial profiles scale with r/x (or y/x) so that a single set of measurements or a correlation can be generalized to other Reynolds numbers in the same geometry [1, 2]. Although some interesting patterns have been observed in wall-bounded flows [3-8], complete and succinct self-similarity structure as found in jet flows had not yet been discovered. A recent work, however, reveals that if one looks underneath the exterior into the "dissipation" structure, an order appears out of the disarray, in which the normal Reynolds stresses ($u'^2$ and $v'^2$) are asymmetrically scalable [9]. For instance, a self-similar pattern emerges in the $u'^2$-gradient structure of channel flows, as shown in Figure 1. It is the dissipation ($du'^2/dy+$) or the gradient profiles that scale with the Reynolds number in an asymmetrical but ordered manner. This bears some implications toward the internal organization processes in turbulence as predicated by the global constraints and transport dynamics, and also renders possible computations of the full flow structure starting from a reference profile [9, 10].

A similar self-repeating sequence was observed for $v'^2$ in channel flows, but at the second gradient level [9], as discussed later. The question of applicability in related flow geometries immediately arises from these observations. In this work, we would like to explore the possible link, and compare the scaling characteristics in wall-bounded flows, including flow over a flat plate. We shall also extend the scaling to the Reynolds shear stress, $u'v'$. Similar to a preceding monograph [9], implications toward a unified turbulence structure are contemplated upon based on this asymmetrical order.

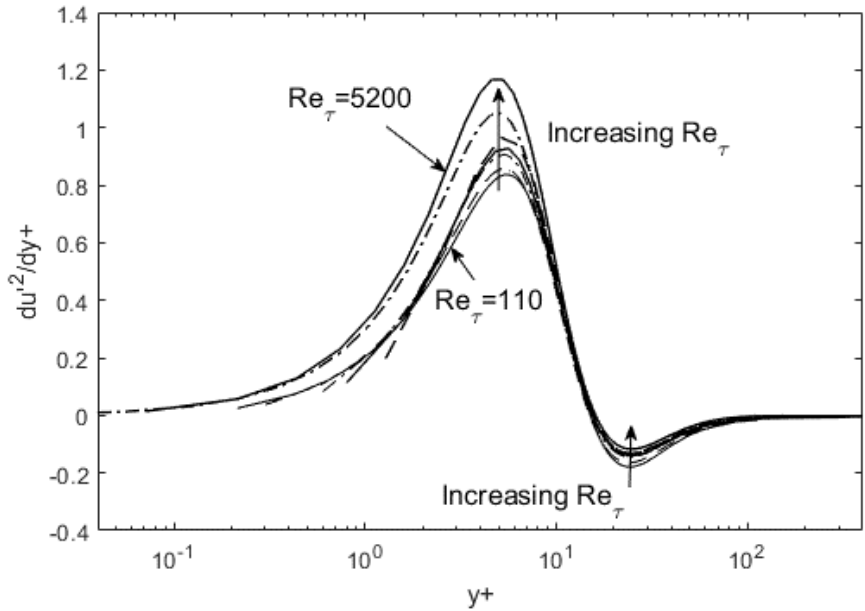

**Figure 1. The dissipation scaling of du'²/dy+. The near-wall peaks ascend, and the negative gradients reduce in magnitude with increasing Reynolds number. DNS data of Iwamoto et al. [11] and Graham et al. [12] are used.**

## DISSIPATION SCALING

Let us begin with the streamwise turbulence kinetic energy, $u'^2$, already seen above in Figure 1. The data for turbulence channel flows are from DNS work of Graham et al. [12], while zero-pressure-gradient boundary layer flow data by Spalart [13] are added in Figure 2. In this work, $u'^2$ and other turbulence variables are implicitly Reynolds-averaged, and also normalized by the friction velocity squared, $u_\tau{}^2$. For example, $<u'^{2+}> = <u'^2>/u_\tau{}^2$, but written as $u'^2$, to abbreviate the notation. In Figure 2, we can see that the scaling observed in channel flows is translatable to boundary-layer flows as well. Overall, they exhibit similar features such as the fixed peak location (zero-crossing in du'²/dy+ plots) in inner coordinates (y+) and increasing steepness with the Reynolds number, followed by a gradual descent toward the outer boundary condition. Figure 2 shows that these characteristics can be surmised or scaled in the dissipation space (d/dy+). The primary peak in du′²/dy+ (the maximum slope in u′²) ascends, and the negative undulations (decay rate) decrease in amplitude. Thus, the scaling is asymmetrical. The proportionality constant monotonically increases for the near-wall segment, while it decreases in the aft side (y+>15). The progression of the peak height in du′²/dy+ is a function of the Reynolds number, with a different proportionality between the first positive and the second negative. In all of these regards, the contours of the gradient profiles exhibit mutually inter-changeable characteristics between channel (CF) and flat-plate (FP) flows, except for having different proportionality constants with respect to the Reynolds number and subtle differences in the boundary conditions. For example, u′² → 0 in flows over a flat plate, and there may be a curvature difference at the outer boundary between the two geometries, which are not easily observable in the resolution of the data plotted in Figure 2.

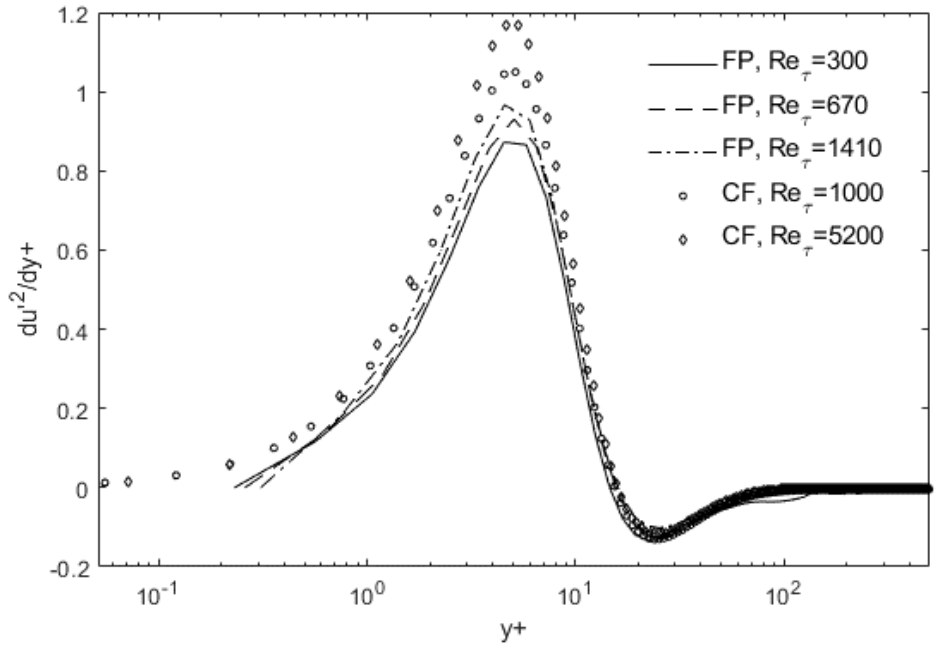

**Figure 2. Combined scaling of du′²/dy+ for flat-plate (FP) and channel flows (CF). The near-wall peaks ascend, while the negative undulations reduce in magnitude with increasing Reynolds number. DNS data of Spalart [13] for FP and Graham et al. [12] for CF are used.**

What about the "outer peak" appearing at high Reynolds numbers [14] or in adverse pressure gradient (APG) flows [15]? The second (negative) perturbation in the gradient in Figure 2 is small in magnitude in comparison to the first, so that any structural features beyond y+ ~ 100 are not easily discernible at these Reynolds numbers. However, flattening of $u'^2$ profiles gradients after the initial peak is observed at moderately high Reynolds numbers [9]. We can visualize from the data [14] that this flattening further develops into a secondary peak. The driving mechanism for the compaction of $u'^2$ profiles toward the wall was thermodynamically attributed to requisite elevation in dissipation, or total integrated $(du'^2/dy)^2$, up to the limit imposed by the viscosity at high Reynolds numbers [9]. This skewing in the $u'^2$ profiles indeed leads to a linear increase as a function of $Re_\tau$ in total integrated dissipation [9]. At yet higher Reynolds numbers, the wall boundary conditions start to impose more stringent viscous suppression so that wall compaction is not as feasible, and the $u'^2$-gradient needs to be produced elsewhere, i.e. a secondary peak far from the wall (y+ = 100 ~ 1000) as observed experimentally and in DNS [14]. It would only take a minute positive/negative gradient, a shade over zero, near y+ of 100 to produce this mid-layer "bulge", as shown in the Appendix. This is akin to a second- or third-harmonic generation in waves to increase the energy content, or in this case dissipation. Judging from the jagged peaks and undershoots at some locations in Figures 2 and 3, we anticipate that high-resolution data would be needed in measuring this fine-tuning in gradient structure at high Reynolds numbers. An exhibition of this harmonic or wave structure leading to the secondary peak in $u'^2$, based on a hypothetical sinusoidal function, is briefly presented in the Appendix, while a more detailed analysis on this awaits in a succeeding study.

A seemingly similar gradient structure is observed for $v'^2$, but at the second order, $d^2v'^2/dy+^2$. In a related work [16], we asserted that $u'^2$ and $v'^2$ transport is governed by alternative expressions of energy and momentum balance, respectively. We discuss the inter-dynamics of these fluctuation components later as well, but a conjecture can be offered wherein $v'^2$ spatial distribution is predicated by the momentum fluxes, the diffusion of which goes as $d^2v'^2/dy+^2$. This induces self-similarity at the second-gradient level. Also, the negative segments beyond the zero-crossing point progress minutely in the downward direction with increasing Reynolds number, an opposite trend from $du'^2/dy+$. Otherwise, the flat plate and channel flow profiles again are interchangeably self-similar at the second-gradient level, as visualizable in Figure 3. Again, recovery of the full $v'^2$ structure from the second-gradients would necessitate stipulations of outer boundary conditions specific to the flow.

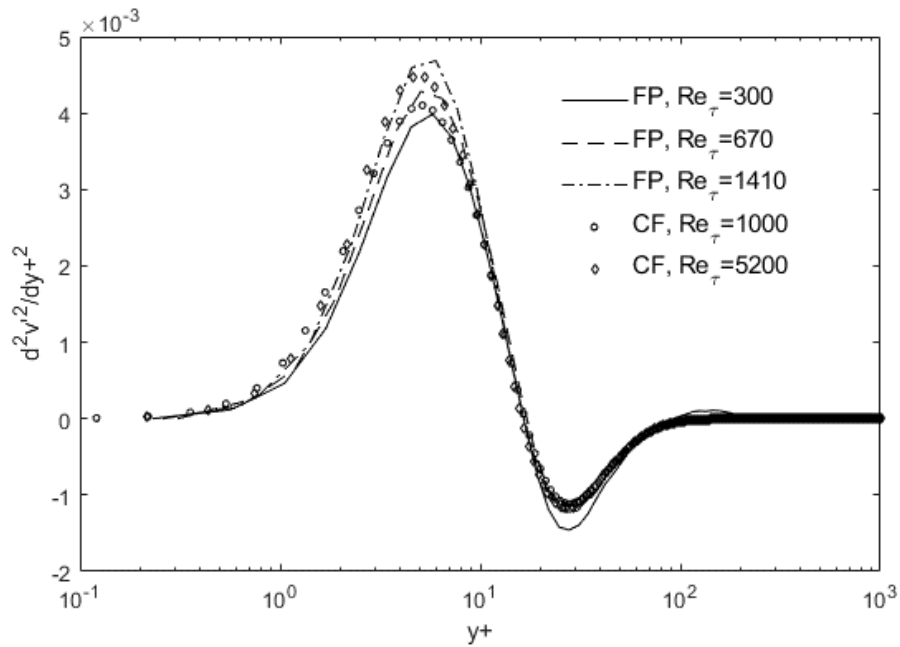

**Figure 3. The dissipation scaling of $d^2v'^2/dy+^2$. The near-wall peaks ascend, and the negative undulations increase in magnitude with increasing Reynolds number. DNS data of Spalart [13] for FP and Graham et al. [12] for CF are used.**

The second-gradient structure for the Reynolds shear stress exhibits a similar asymmetrical order, except inverted from $d^2v'^2/dy+^2$. There is a small off-set in y+ coordinates between the lines (flat plate) and symbols (channel), but this can be corrected with more accurate estimation of the friction velocity. A notable pattern is that the height of $d^2u'v'^2/dy+^2$ peaks changes very slowly with respect to the Reynolds number, less than the primary peak changes for prior variables. Nonetheless, the self-similar patterns observed in Figures 2-4 lead to easily reproducible profiles for the gradients of $u'^2$, $v'^2$ and $u'v'$ based on a common template for each variable at a reference Reynolds. Some boundary condition corrections at the outer edge are needed as discussed next.

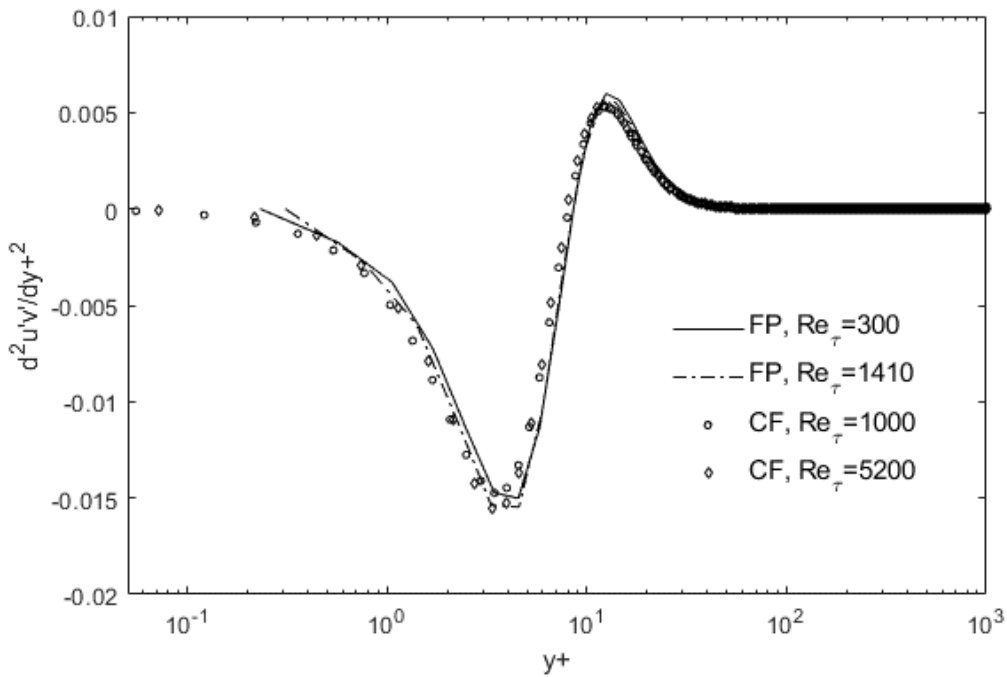

**Figure 4. The dissipation scaling of $d^2u'v'/dy+^2$. DNS data of Spalart [13] for FP and Graham et al. [12] for CF are used.**

The self-similarity patterns in the gradient structure of the Reynolds stresses can be summarized as follows ($a_{ij}$ are proportionality constants for $(df/dy+)_o$ at a reference Reynolds number):

$$du'^2/dy+ = \quad a_{11}(du'^2/dy+)_o \text{ for } y+ < y+_1 \qquad\qquad (1)$$
$$a_{12}(du'^2/dy+)_o \text{ for } y+ > y+_1$$

$a_{11}$ increases, $a_{12}$ decreases with $Re_\tau$. $y+_o$ is the zero-crossing point in $du'^2/dy+$, or the peak location in the $u'^2$ profile.

$$d^2v'^2/dy+^2 = \quad a_{21}(d^2v'^2/dy+^2)_o \text{ for } y+ < y+_2 \qquad\qquad (2)$$
$$a_{22}(d^2v'^2/dy+^2)_o \text{ for } y+ > y+_2$$

$a_{21}$ increases, $a_{22}$ increases with $Re_\tau$. $y+_2$ is the zero-crossing point in $d^2v'^2/dy+^2$, or the inflection point in the $v'^2$ profile.

$$d^2u'v'/dy+^2 = \quad a_{31}(d^2u'v'/dy+^2)_o \text{ for } y+ < y+_3 \qquad\qquad (3)$$
$$a_{32}(d^2u'v'/dy+^2)_o \text{ for } y+ > y+_3$$

$a_{31}$ increases, $a_{32}$ increases with $Re_\tau$. $y+_3$ is the zero-crossing point in $du'v'^2/dy+^2$, or the inflection point in the $u'v'$ profile.

Let us see how reconstructions of gradient profiles would work from the above scaling order, which surmises that simple multiplicative expansions of the near-wall and aft sides separately from a baseline profile would lead to recovery of the full structure at other Reynolds numbers. As one example, we take a $u'^2$ profile from DNS data [11] for channel flow at $Re_\tau = 300$ (arbitrarily chosen), numerically differentiate to obtain $du'^2/dy+$, then perform the scaling operation. In order to check the interchangeability in the self-similarity, we apply the scaled (expanded) dissipation profiles to experimental data for boundary-layer flows over a flat plate [17]. Since $u'^2$ data from experiment are not accurately differentiable, we integrate back the scaled $du'^2/dy+$ profiles for direct comparison, as shown in Figure 5. The exact progressions of the constants, $a_{11}$ and $a_{12}$, are currently being cataloged based on expanded data sets. At this point, however, knowing the monotonicity in the Reynolds number dependence, we adjust these constants, and only these, with the data to generate the final outcome. From the results of this exercise shown in Figure 5, we can conclude that through a simple scaling procedure the $u'^2$ profiles can be constructed at any Reynolds number, starting from a single reference template of $du'^2/dy+$, within [9] and across different wall-bounded flows. Even the exotic peak shapes at higher Reynolds numbers in Figure 5 are merely due to the progression in the gradient structure. A large gradient followed by a shallow undulation will result in a sharpened peak of $u'^2$ close to the wall with an abrupt bend toward a gradual descent toward the outer boundary. One limiting factor is that y+ range increases with the Reynolds numbers so that starting from low-$Re_\tau$ template cuts short the extent to which the profiles can be recovered far from the wall. However, this can mostly be remedied by enforcing and extrapolating to the outer boundary conditions, up to the second order. The boundary condition correction is particularly essential when one goes across from CF and FP or vice versa, since there are subtle differences in the $u'^2$ and its curvature at the centerline and outermost location.

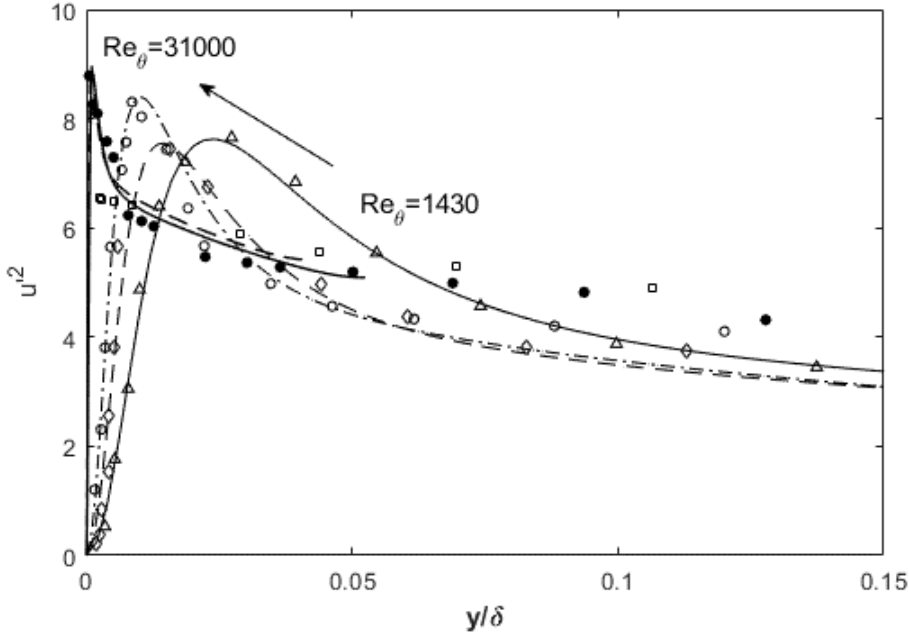

**Figure 5. Reconstruction of flat-plate u′² from a channel-flow template at Re_τ = 300. The template is taken from DNS data of Iwamoto et al. [11], while experimental data are from deGraaf and Eaton [17].**

Similar reproducibility in the Reynolds shear stress (u′v′) is demonstrated in Figure 6, wherein double integration is needed to go from the scaled second-gradient structure (Figure 4) of channel flow to boundary-layer flow. For higher resolution, we use the DNS data [12, 13] this time for validation, which again shows the cross-applicability (channel ←→ flat plate) in dissipation scaling in Figure 6. As noted above, the different outer boundary conditions induce deviations far from the wall, which again can be corrected by applying the outer constraints up to the second order. The end point, slope and its curvature all must match at the outer edge of the flow. Other than this deviation toward the ambient boundary and any possible numerical errors during differentiation/integration in Figure 6, seemingly inexplicable Reynolds shear stress obeys the same scaling principle as other root turbulence variables. This can be leveraged to compute it at an arbitrary Reynolds number, leading to computability of the mean flow structure [9]. Of equal or more significance is that u'v' or its gradients follow a sensible and predictable transport dynamics as shown next.

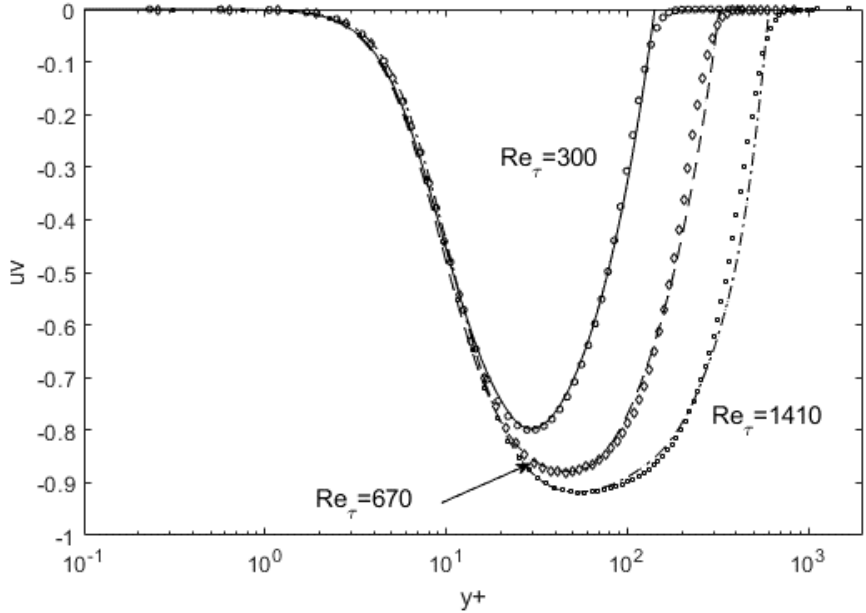

**Figure 6. The Reynolds shear stress reconstructed using Eq. 1 and numerical integration (lines), based on a channel-flow template at Re$_\tau$ = 1000 [12]. DNS data (symbols) are for boundary-layer flow over a flat plate from Spalart [13].**

**FLUX MECHANICS OF THE TURBULENCE STRUCTURE**

The mechanics of the above scaled profiles can be analyzed through the recently formulated transport equations [16, 18] for the Reynolds stress tensor components.

u' momentum transport:

$$\frac{d(u'v')}{dy} = -C_{11}U\frac{d(u'^2)}{dy} + C_{12}U\frac{dv'^2}{dy} + C_{13}\frac{d^2u'}{dy^2} \tag{4}$$

v′ momentum transport:

$$\frac{d(v'^2)}{dy} = -C_{21}U\frac{d(u'v')}{dy} + C_{22}U\frac{dv'^2}{dy} + C_{23}\frac{d^2 v_{rms}}{dy^2} \tag{5a}$$

Alternatively,

$$\frac{d(v'^2)}{dy} = \frac{-C_{21}U\frac{d(u'v')}{dy} + C_{23}\frac{d^2 v_{rms}}{dy^2}}{1 - C_{22}U} \tag{5b}$$

u′² transport:

$$\frac{d(u'^3)}{dy} = -C_{31}\frac{1}{U}\frac{d(u'v'\cdot u')}{dy} + C_{32}\frac{1}{U}\frac{d(v'\cdot u'v')}{dy} + C_{33}\frac{1}{U}\left(\frac{du'}{dy}\right)^2 \tag{6}$$

In these equations, the fluctuating terms, u′v′, v′², etc., are all Reynolds-averaged. The concepts and hypotheses contained in Eqs. 4-6, and their efficacy in prescribing the Reynolds stress tensor, are described in in the Appendix and Lee [16, 18], respectively. Upon a brief examination of the flux terms, we can see that Eqs. 4 and 5 are momentum-conserving, while Eq. 6 is an expression of energy balance. The transport equations involve gradients up to the second order, as diffusion/dissipation terms, which may have some implications toward the scaling behavior [9].

The progressions in previous shown gradients (Figs. 2-4) exhibited self-similarity within each of the variables, but a discernible pattern was also visible across them as well.

For example, $d^2u'v'/dy+^2$ has an inverted shape of $du'^2/dy+$, which in turn has a similar appearance with $d^2v'/dy+^2$ except with an asymmetrical development on the aft side of the zero-crossing. The inter-dynamics between $u'v'$ and $u'^2$ can be anticipated from Eq. 4, which shows that the Reynolds shear stress gradient is to the leading order proportional to negative (minus) of $du'^2/dy+$, with U acting as a modulator [16, 18]. This will result in an inversion and thus folding with the $d^2u'v'/dy+^2$ as shown in Figure 7, where once-more differentiated terms of both sides of Eq. 4 are plotted and again compared with DNS data of Graham et al. [12]. On the aft side, the structure is modified by the pressure ($C_{12}$) term in Eq. 4. Combining both terms, along with the viscous term reproduces the observed scalable structure (second gradient) for the Reynolds shear stress, to a level of accuracy rarely seen in unphysical modeling. Once $u'v'$ is known through integrations of $d^2u'v'/dy+^2$, then the mean velocities can be computed through the respective Reynolds-averaged Navier-Stokes equations.

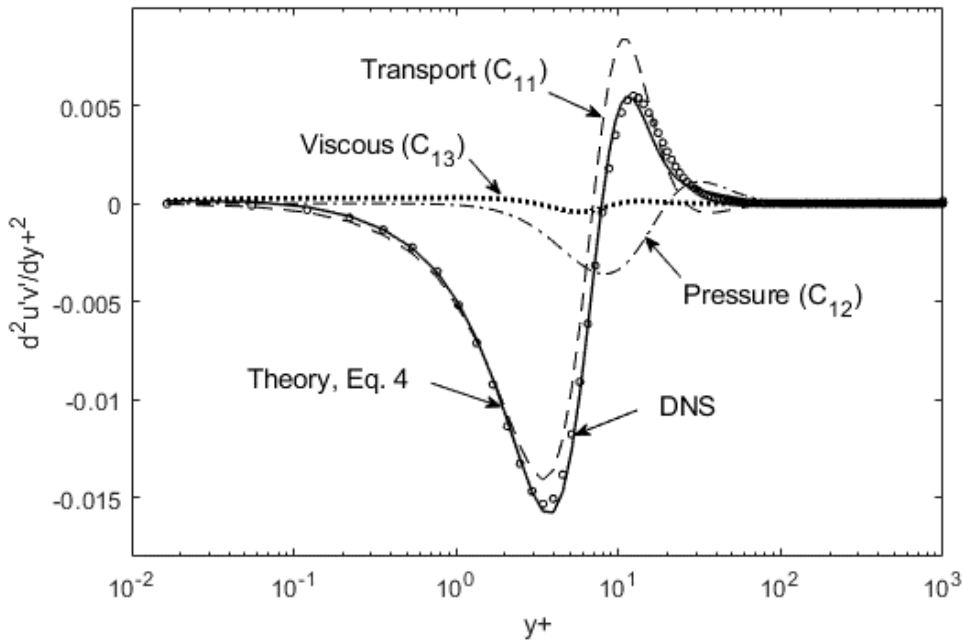

**Figure 7. Theoretical prediction of the observed dissipation structure for the Reynolds shear stress, $d^2u'v'/dy+^2$. DNS data of Graham et al. [12] are used for comparison.**

Similar flux balances between $u'^2$, $v'^2$ and $u'v'$ can be checked through Eq. 5 or 6. Since the first gradient is slightly easier and involves less numerical errors, we shall examine the $u'^2$ gradient structure through Eq. 6. Figure 8 shows that the sum of the transport ($C_{31}$) and pressure work ($C_{32}$) terms in Eq. 6 add up nearly perfectly to the observed $du'^2/dy+$. The zero-crossing point is the location of observed $u'^2$ peak(s) at y+ ~ 15.4 in channel flows. The viscous dissipation ($C_{33}$) term may slightly modify the structure near the wall, but at high Reynolds numbers it appears to be negligible relative to other two terms in Eq. 6. Figure 8 proves that just a simple energy balance of Eq. 6 is the physical origin of the gradient structure, which upon integration returns the sharply-peaked $u'^2$ profiles. The peaks near the wall arises due to the increasing $du'^2/dy+$ with $Re_\tau$. Then, aft of the zero-crossing point (y+ ~ 15.4), the negative slopes on the reverse side decreases in amplitude with the Reynolds number, which inevitably produces the bend in the downward sides of $u'^2$ profiles, e.g. deGraff and Eaton [17] data in Figure 5. This pattern for $u'^2$, seemingly not traceable with any known algebraic functions, can be faithfully tracked in the "gradient space", all based on a transport equation for streamwise turbulence kinetic energy (Eq. 6). On a side note, there may be not suitable function fits for the complete profile in the physical space (y+), but interestingly Weibull function appears to follow the near-wall $du'^2/dy+$ quite closely up to the peak location (see Appendix). This offers a possibility for the usage of piecewise continuous functions for representation of the dissipation structure, so long as subtle differences in the outer boundary conditions are accounted for channel and flat plate flows.

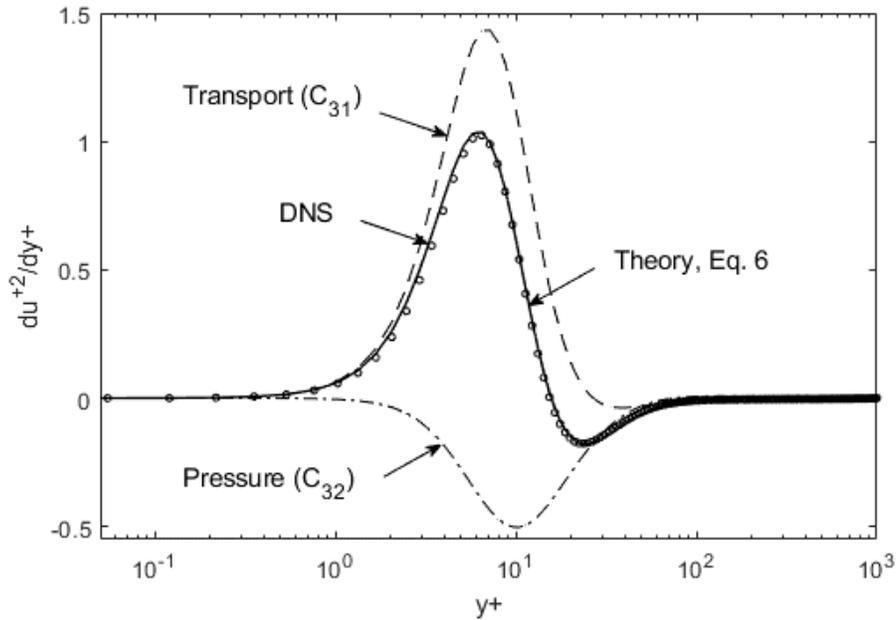

**Figure 8. The energy flux budget for u′²-gradient. Again, DNS data [12] is used for confirmation of Eq. 6.**

**CONCLUDING REMARKS**

The dissipation scaling in wall-bounded turbulent flows, based on the above observations, appears to be complete and uniformly applicable, save for the adjustments for the outer boundary conditions. The self-similar characteristics mean that a single solution can be extended to other Reynolds numbers through simple reconstructive operations. The inter-dynamics of the root turbulence variables are also demonstrated through the transport equations (Eqs. 4-6) for the Reynolds stress. Further confirmations await for the secondary structure in u′² profiles at high Reynolds numbers. Even in this regard a hypothetical second harmonic dissipation "wave" regenerates this feature. Combined with the flux mechanics of Eqs. 4-6, the asymmetrical order embodied in wall-bounded flows renders possible understanding and reconstructions of the full turbulence structure.

## APPENDIX:  Derivation of Transport Equations (Eqs. 4-6)

Figure 9(a) shows transport of momentum (v′) and energy (u′²), while Figure 9(b) illustrates the displacement transform discussed below.  From Figure 9(a), the transport equations for v′ (lateral turbulence fluctuation momentum) and u′² (longitudinal turbulence kinetic energy) follow the same logic as the u′-momentum [16] in that the net momentum (or energy) flux is balanced by the pressure and viscous force (or work) terms for the control volume moving at the local mean velocity.  This leads to the following v′-momentum (Eq. 4) and u′²-energy equation (Eq. 6):

$$\frac{d(u'v')}{dx} + \frac{d(v'^2)}{dy} = -\frac{1}{\rho}\frac{dP}{dy} + \nu\frac{d^2 v_{rms}}{dx^2} \tag{7}$$

$$\frac{d(u'^2 u')}{dx} + \frac{d(u'^2 v')}{dy} = \frac{d(Pu')}{dx} + 2\nu\left(\frac{du'}{dy}\right)^2 \tag{8}$$

The differentiated terms (in the parentheses) are again Reynolds-averaged, e.g. u′v′ implicitly is <u′v′>, etc.  Since the relative velocities (U, V) are zero for the moving control volume, this will set UdU/dx and VdU/dy terms to zero and thus do not appear in Eqs. 4 and 5.  For the mean shear stress terms, such as μd²U/dy², which does not necessarily go to zero under Galilean transform, we can take the view that the turbulent shear is much larger than the laminar viscous stress, du′rms/dy >> dU/dy.

Now, we introduce the d/dx to d/dy transform, which is applied in all the transport equations.  This is based on the displacement concept as shown in Figure 9(b), where a control volume moving in the boundary layer would see a lateral shift in the vertical profile [14, 15].  For channel flows, an extension of this transform is required, as analyzed in the Lee (2020).  We will outline this transform as the first of the several unique hypotheses embodied in this formalism, some of which depart from traditional turbulence algebra.

Due to the displacement effect, d/dx is converted to a d/dy term (Eq. 9) with the mean velocity acting as the modulator [16]. As illustrated in Figure 9(b), the boundary layer displacement effectively leads to a gradient in the vertical direction for a translating control volume, and this effect will be more pronounced more large U. For channel flows bounded by walls, there are no displacement effects; however, a "probe transform" bearing the same functional result has been analyzed in Ref. 16.

$$\frac{d}{dx} \rightarrow \pm C_{ij} U \frac{d}{dy} \qquad (9)$$

$C_{ij}$ is a constant with the unit, $\sim 1/U_{ref}$. The +/- sign depends on the flow geometry, i.e. the direction of displacement relative to the reference point. The dreaded triple correlations in $u^2$-transport equation (Eq. 4) are treated in a similar manner as validated in the previous works [16].

$$\frac{d(u'^2 v')}{dy} \approx \frac{d(u' \cdot u' v')}{dy}, \ \frac{d(u' v'^2)}{dy} \approx \frac{d(v' \cdot u' v')}{dy}, etc. \qquad (10)$$

This simplistic and expedient differentiation rule decomposes the triple product in terms of existing variables, and leads to unexpectedly accurate results for $u'^2$ and $v'^2$ profiles, as shown above. The key conceptual distinction in the current formalism, relative to the Eulerian, is that these products represent flux terms, and thus should be treated accordingly. The second of the triple product, $u'v'^2$, appears in Eqs. 6 and 8 when the pressure is written as below.

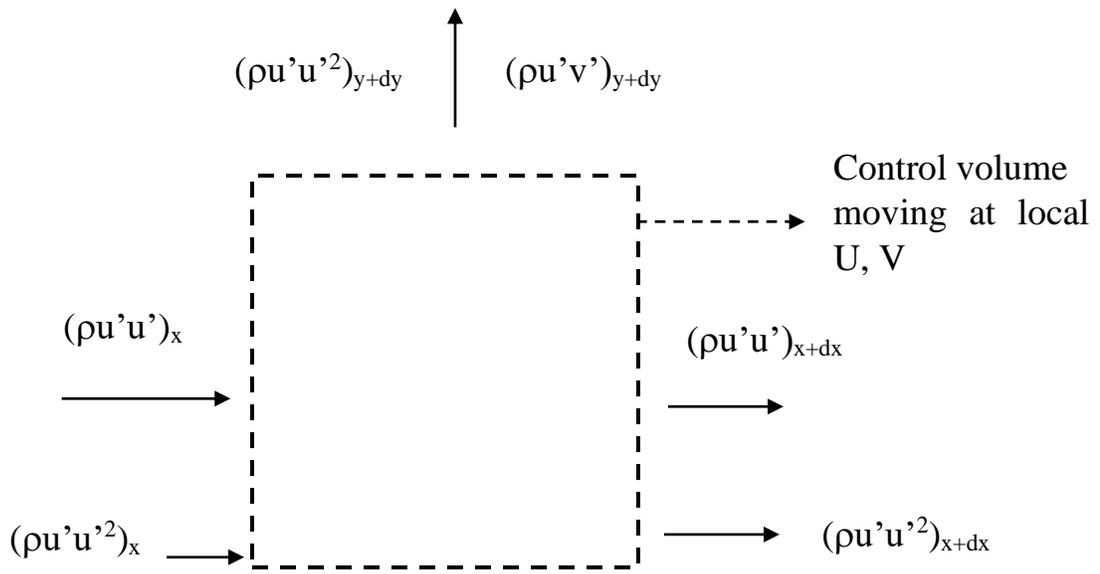

(a)

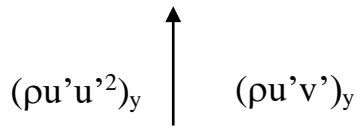

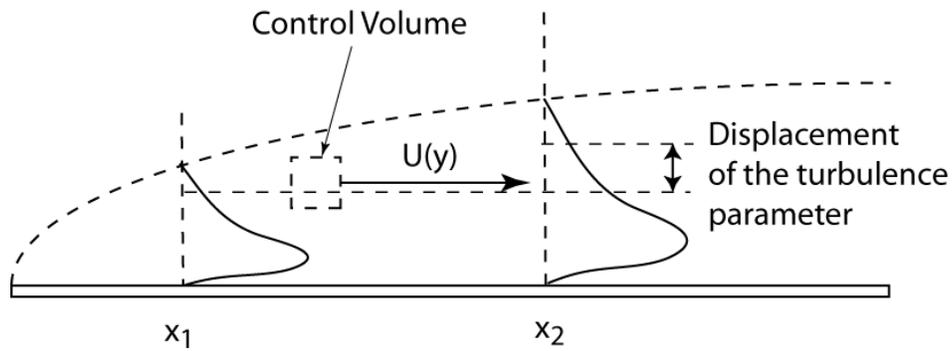

(b)

**Figure 9. Schematics of the dynamics contained in the Lagrangian turbulence transport: (a) y-momentum (v') and longitudinal kinetic energy balance (u'²) following a control volume, moving at the local mean velocity; and (b) the displacement concept leading to d/dx → d/dy transform.**

$$P \approx -\rho v'^2 \qquad (11)$$

For channel flows, this relation is exact [2], derivable from the y-component of the RANS, and we apply it as an approximation for boundary layer flows with and without adverse pressure gradient. The pressure expression (Eq. 11) leads to d(Pu')/dy → d(u'v'²)/dy in Eq. 8, which is converted to d(v'(u'v'))/dy via Eq. 10. The last hypothesis is that the pressure work done on the moving control volume is mainly due to the mean pressure-u' combination:

$$Pressure\ work \approx \frac{d\left(Pu'\right)}{dx} \qquad (12)$$

These are evidently new and unique concepts, nonetheless they lead to a complete and symmetric set of transport equations (Eqs. 4-6) for u'v', u'² and v'², and agree quite well with DNS results as shown above and in previous works [9, 10, 16].

**Function Approximations**

It would be of much convenience if the dissipation structure could be approximated with piece-wise differentiable functions. Then, they can serve as a reference template. For a reason not yet explored, the near-wall segment of the du'²/dy+ profile follows the Weibull distribution function quite closely (Figure 10), so that a simple "stretching" of this form can re-generate the spatial distribution at other Reynolds numbers. On the aft-side, chi-square function is a better fit, but with less accuracy. Function approximations for other variables require further attention.

To examine a possible root cause of the secondary undulation in $u'^2$ profiles observed at high Reynolds numbers [14], we can hypothesize a second-harmonic generation and check its effect. Although high-resolution data for this secondary structure have not yet been accessed or analyzed, we can still try with a synthesized sinusoidal function which dilates in the y+ coordinate and also undergoes decay in magnitude (shown as a dilating sinusoid in Figure 10). Let us see what follows from numerical integration of this hypothetical dilating sinusoid, along with the Weibull function for the near-wall region, as shown in Figure 11. The sinusoidal function is exaggerated in Figure 10 to render visibility over other data. It only requires a minute positive/negative slopes in this y+ range to cause flattening or bulging in the mid-layer, as shown in Figure 11. No comparison with data is attempted at this point, but this formalism is being advanced toward that end.

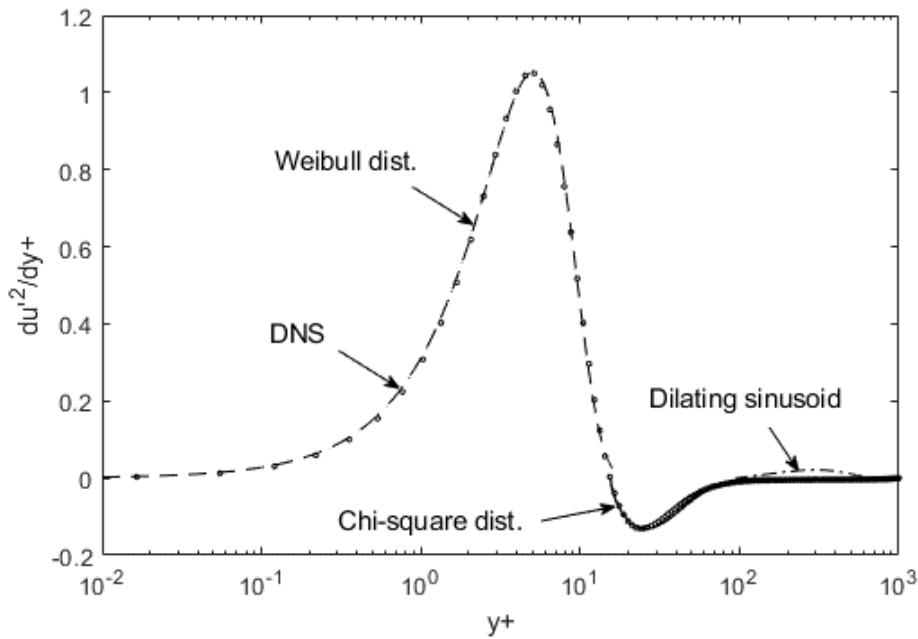

**Figure 10. Weibull and chi-square distribution functions and $u'^2$-gradient. A hypothetical sinusoid is added, which produces the secondary peak in the $u'^2$ profile shown next.**

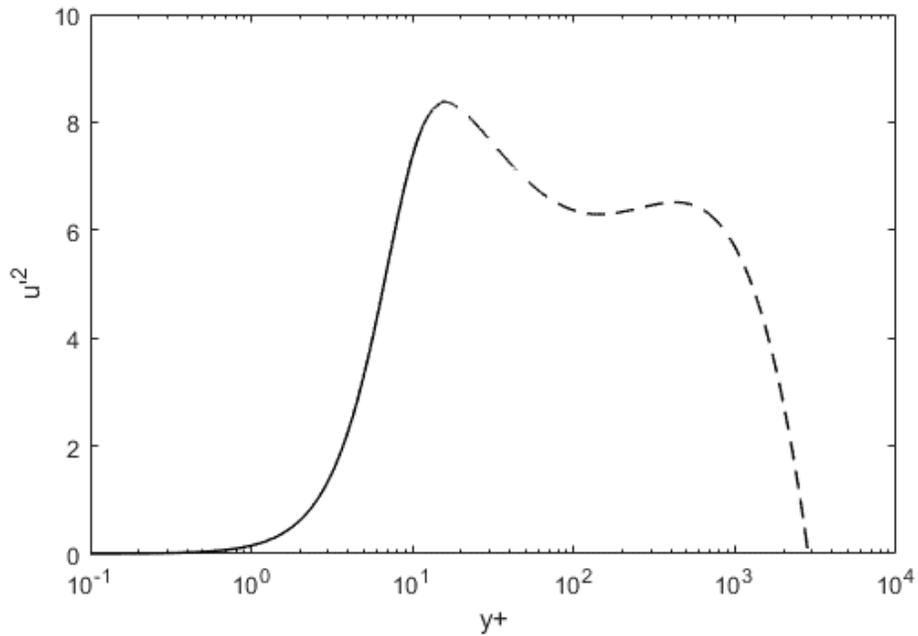

**Figure 11. The primary peak near the wall is from the Weibull function, while the secondary structure is derived from numerical integration of the hypothetical sinusoid for $du'^2/dy+$ beyond $y+ = 100$.**